# From Artifacts to Outcomes: Comparison of HMD VR, Desktop, and Slides Lectures for Food Microbiology Laboratory Instruction

Fei Xue
University of California, Davis
Davis, CA, USA
feixue@ucdavis.edu

Rongchen Guo
University of Texas at Austin
Austin, TX, USA
rongchen_guo@utexas.edu

Siyuan Yao
University of Notre Dame
Notre Dame, IN, USA
syao2@nd.edu

Luxin Wang
University of California, Davis
Davis, CA, USA
lxwang@ucdavis.edu

Kwan-Liu Ma
University of California, Davis
Davis, CA, USA
ma@cs.ucdavis.edu

## ABSTRACT
Despite the value of VR (Virtual Reality) for educational purposes, the instructional power of VR in Biology Laboratory education remains under-explored. Laboratory lectures can be challenging due to students' low motivation to learn abstract scientific concepts and low retention rate. Therefore, we designed a VR-based lecture on fermentation and compared its effectiveness with lectures using PowerPoint slides and a desktop application. Grounded in the theory of distributed cognition and motivational theories, our study examined how learning happens in each condition from students' learning outcomes, behaviors, and perceptions. Our result indicates that VR facilitates students' long-term retention to learn by cultivating their longer visual attention and fostering a higher sense of immersion, though students' short-term retention remains the same across all conditions. This study extends current research on VR studies by identifying the characteristics of each teaching artifact and providing design implications for integrating VR technology into higher education.

## CCS CONCEPTS

• **Human-centered computing**; • **Human computer interaction (HCI)**; • **Interaction paradigms**; • **Virtual reality**;

## KEYWORDS

Virtual Reality, Educational Technology, Laboratory Instruction, Immersive Visualization Design; Learning Theories

## 1 INTRODUCTION

In recent years, scholars have investigated how VR enhances learning in various disciplines, including physics [86, 90, 110], medicine [70, 83], chemistry [19, 29], engineering [27, 59], linguistics [10, 28, 105], and many others. However, a comprehensive investigation into the instructional effectiveness of VR on undergraduate biology laboratory education has yet to be done. For example, Food Microbiology is a critical scientific study of microbes, which play a vital role in all life on earth. In a Food Microbiology Laboratory course, students are trained in laboratory methods used in the microbiological analysis of foods. Microorganisms play significant roles in the production, manufacturing, preparation, and storage of food items. However, learning food microbiology concepts and microbiology laboratory courses can be challenging for some students due to low retention of scientific literacy skills [3, 13] and lack of sustained attention to conceptual learning [72].

As a result, educators and researchers have been using emerging technologies to address this concern. Both visualization and VR have been integrated into technology-based learning approaches to facilitate authentic scientific knowledge transmission [86] and promote students' learning interests [83, 102]. Based on the theory of distributed cognition [42], educational technologies can serve as an extension of the human mind, supporting an individual's sense-making. As one of the most promising tools across various technologies, VR encourages learning by isolating users from the physical world and facilitating high levels of immersion and interactivity in the virtual world [21, 37]. However, how learning happens throughout these interactive haptic simulations (e.g., sensor gloves and hand controllers) and augmented visual perceptions (e.g., 360-degree visual displays) is still unclear. Most existing VR studies assess learning outcomes using error rate [50, 106], task completion time [8, 79, 99, 109], and short questionnaires [10, 108, 112] in lab setting. Although some of these findings show that VR can provide immediate learning gains, they do not identify which factors from such computer-mediated environments to one's individual cognition, contribute to improved learning attainments during the VR experience.

Moreover, while research has demonstrated positive learning experiences in immersive environments, scholars have not yet investigated the impact of increased motivation on students' learning performance in microbiology laboratory courses. PowerPoint slides are commonly used in current Food Microbiology laboratory instruction in undergraduate education, but it's hard to maintain students' interest without effective teacher-student interaction [6, 12]. As mentioned earlier, teaching microbiology can be challenging due to students' low motivation to learn abstract scientific concepts and low retention rates. To address these existing learning barriers in microbiology undergraduate education, technology integration in classrooms should ease teachers' burden, improve academic attainment and cultivate students' motivation to learn. Last but not least, a theoretical framework derived from learning science to guide designing immersive visualization is still lacking [89]. There is an urgent need to examine the effectiveness of theory-driven learning approaches in VR due to the current lack of empirical evidence in this field. Also, educational researchers can implement immersive visualization based on appropriate pedagogical guidance to better understand, discover, and analyze learning behaviors.

In our work, we developed a VR application as a self-directed learning approach to study how students memorize Food Microbiology lectures. Using this tool, students' learning processes, outcomes, motivation, and perceptions were investigated. The self-directed VR lecture was adapted to two conventional learning materials - PowerPoint slides and a desktop application. This study utilized achievement tests (retention), self-report measures (perceived immersion and motivation), and observational measures (visual attention) to comprehensively examine students' learning process as well as the process of human-computer interaction throughout different instructional artifacts. In addition, interviews were conducted to understand students' perceptions of learning with each artifact. Our research design is guided by the theory of distributed cognition and motivational theories.

This study builds on existing HCI research about educational VR and learning theories. With the rapid development of emerging technologies, a systematic investigation into students' learning experiences during and after VR lessons, coupled with students' perceptions of their engagement with VR, will provide evidence-based insights and directions for future research. Specifically, this study contributes to the following areas: (1) offering a theoretical framework that can be used to guide the design and evaluation of learning with technologies, (2) evaluating the effects of HMD VR, desktop application, and PowerPoint slides, and finding significant differences in students' long-term retention, learning motivation, immersion, and visual attention, (3) examining the learning process and learner behaviors through their immersive and non-immersive experiences, (4) identifying students' learning strategies using different instructional artifacts, and (5) summarizing design and research implications of selected instructional artifacts usage in higher education. Two questions were asked:

- **RQ1**. How does learning occur among different instructional artifacts (HMD VR vs. desktop vs. slides) in self-directed learning?
  - **RQ1.a** What are the differences in the perceived immersion, visual attention, short-term retention, long-term retention, and motivation among HMD VR, desktop, and slides?
  - **RQ1.b** How do students' visual attention, perceived immersion, and motivation contribute to their cognitive processing when learning from different instructional artifacts?
- **RQ2**. How do students perceive their learning experiences from the three instructional artifacts?

## 2 RELATED WORK

In this section, we present a brief review of the previous studies in general to understand learning in Food Microbiology Laboratory lecture through VR.

### 2.1 VR and Education

Virtual environments have been considered powerful for supporting learning for decades [55, 63, 70, 82, 83, 101]. A 3D desktop display can simulate non-immersive virtual environments when users use a mouse or keyboard to manipulate (e.g., rotate and drag) objects on computer screens [45, 54, 63]. Users can perceive a believable experience similar to real life while staying connected with their physical surroundings [26, 45]. For example, Mei and Sheng [70]used a desktop-based virtual hospital system to test medical students' practice of human organ anatomy. Their findings suggested that the virtual environments within the desktop display can stimulate learner motivation towards medicine while helping learners improve their clinical skills. Nowadays, with more affordable VR devices being developed, Head-mounted displays (HMD) have been widely used for educational aims as they provide fully immersive experiences to make users feel an illusion that they are completely isolated from their physical world [45, 67]. Prior work has shown HMD VR may scaffold users' situated learning experience [18, 98] and emotional involvement [44, 94, 109] by enabling learners to enter, navigate, manipulate, and interact with virtual three-dimensional objects. Hence, comparing immersive and non-immersive VR has become an interest in the field. For instance, Oberdörfer et al. [82] explored the possibility of using HMD VR to leverage participants' gamified knowledge and demonstrated the value of HMD VR over desktop versions. On the contrary, Feng [24] found that the desktop version can better support pedestrian route behavior than HMD VR. Though the above work suggested instructional affordances of VR to some extent, studies comparing knowledge-based learning outcomes between VR and other technological tools [24, 82] or VR and conventional teaching methods [101, 113] are still at the exploratory stage. Because most of these works were limited to evaluating immediate learning in a lab setting so their results cannot allow for generalization. Therefore, there is a strong need [45, 62] to systematically assess how VR effects can be maintained or shaped over time compared to other novel technologies (e.g., desktop application) and traditional teaching methods (e.g., PowerPoint slides). Such comparison will allow us to advance the ongoing body of work in educational VR.

### 2.2 Immersion in VR

Engaging with VR allows users to interact with synthetic environments, making them feel as if they have been in a real-life scenario, which is difficult to achieve from traditional "desktop" visualization tools. This realistic mental involvement is mainly due to the unique characteristics of immersion in VR. Studies [9, 81, 83] have discussed several influences of immersion when learning in VR . For example, Chen et al. [9] and Liu et al. [57] showed that providing immersion in VR may promote middle school students' situational interest in the study materials. Furthermore, the immersive nature of VR has been intuitively related to users' satisfaction when simultaneously interacting with virtual objects [78]. As subjective perceptions of pleasance and curiosity, satisfaction and interest can be considered critical psychological factors to reflect one's actual feeling of being motivated.

Therefore, findings from prior research revealed immersion plays an essential role in facilitating users' affective perceptions during their learning process. Understanding if immersion in VR can promote students' learning motivation is vital for VR researchers because it can guide future research design and assessment methods.

## 2.3 Food Microbiology Instruction

Although literature have investigated how VR aids students' academic achievement in many fields [10, 19, 29, 83], examination of VR usage in teaching Food Microbiology is minimal. In Food Microbiology laboratory instruction, both declarative and procedural knowledge is very important for students entering the workforce [87]. However, laboratory lectures are sometimes not offered in the same setting as laboratory experiments. Instead, students have their PowerPoint-based lectures in a traditional classroom and then move to a different location for the laboratory exercise. Sometimes those laboratory exercises can take several days. In this case, longer knowledge retention is critical for students as they must complete precise procedures required for that particular laboratory exercise based on their memory recall [23]. Technology such as desktop games and VR integrates active learning elements into the science curricula and benefits students' critical thinking [19, 55]. However, it is unknown how students perceive or perform differently from such active learning using desktop applications and VR. Our study seeks to provide practical insights that can help instructors to plan and deploy their laboratory lectures.

## 2.4 Visualization Aids Education

Visualization has been widely used for communicating complex data and abstract scientific concepts through augmenting human perception [61, 77], directing visual attention [34, 36], and facilitating comprehension [1, 31, 107]. Prior work has explored the interconnections between visualization design and educational objectives [1, 95]. Compared with traditional desktop-based visualization, a higher graphic rendering in HMD VR can make abstract scientific concepts more accessible and allow exploration from a first-person perspective [8, 48, 66]. For example, Kwon et al. [53] demonstrated that participants tend to answer difficult questions using less time in immersive graph visualization than in the 2D graph version. Mazur et al. [69] found that gestures and verbalization through immersive visualization can support engineering students' general learning articulation.

The above literature indicates an inherent connection between visualization and learning science. This claim aligns with Liu et al. [58]'s argument that an effective visualization should serve as an extended human mind to amplify cognition. Coupled with the immersion nature of VR, visualization linked affordances of augmenting visual awareness, revealing its potential for leveraging cognitive learning.

## 3 THEORETICAL FRAMEWORK

The benefits of VR in education have not yet been fully leveraged or investigated. While educational researchers acknowledge the instructional benefits of VR, most present VR tools are ready-to-use applications instead of personalized learning materials and lack dedicated pedagogical guidelines for application design [46]. These ready-to-use educational VR applications cannot dynamically configure their specialized cognitive properties to extend the human mind to meet specific learning objectives. The theory of distributed cognition (DCog) has been demonstrated to account for the human-computer interaction phenomenon - visual representation is a form of external cognition, and interaction facilitates cognitive tasks [38]. Moreover, motivational theories [15, 74, 83, 100] explain how one's beliefs and values influence their development and achievement. For example, *interest theory* [100] reveals that with proper intervention, students' temporary curiosity about study materials can be transferred to an inherent satisfaction with learning. Such inherent satisfaction denotes one's motivational appeal during the task, from the perspective of *Self-determination theory* [15]. Therefore, taking DCog and motivational theories together, these ideas form the basis for the study design and guide our research hypotheses.

## 3.1 Distributed Cognition (DCog)

First proposed by Hutchins in 1997, distributed cognition holds that cognition is more of an emergent property of interaction than a property of the human mind. Hutchins demonstrated how cognitive tasks of ship navigation involve cooperation between people and the artifacts [7]. In 2000, Hutchins and his colleagues described the idea of a "distributed cognitive system," which is widely applied in HCI research [38]. The central argument is that designers of human-computer interaction systems should consider manipulating graphic representations and interactions as forms of external cognition of humans and that these visual representations and interactions can impact cognitive tasks such as reasoning and thinking. In this respect, learning through a computer-mediated system is an embodied activity across oneself and the world.

Visualization uses computer-supported, symbolic, and metaphorical graphic rendering to represent data and information. Representation and interaction are two major components of information visualization, serving as an external representation of the human mind for carrying out cognitive decisions [42, 58]. How information is processed through interaction between learners and the structures in their environment has been studied in prior work [2, 4]. In VR space, this information processing can be deepened by increased immersion and spatial navigation. A high level of immersion in VR mimics learning material and environments that people are familiar with and encultured with [50]. In addition, virtual spatial navigation triggers interaction between the knowledge domain and the learner, leading to experiential learning [40]. Meanwhile, the interaction is facilitated by learners' visual perception, gestures, and movements in the physical world, and 360-degree virtual simulations amplify learners' visual perceptions. Thus, the cognitive process in VR occurs across its distributed heterogeneous systems that combine embodiment, space, culture, and virtual artifacts. Compared with traditional instructional media such as PowerPoint slides and desktop applications, VR has great potential for supporting an extended cognition derived from the DCog.

Table 1: Comparing HMD VR, Desktop Application, and Slides by Media Feature.

| | Feature | HMD VR | Desktop | Slides |
|---|---|---|---|---|
| **Representation** | Static visuals | ✓ | ✓ | ✓ |
| | Animated visuals | ✓ | ✓ | |
| | 360° visuals and video | ✓ | ✓ | |
| **Interaction** | Physical interaction (haptic) | ✓ | ✓ | ✓ |
| | Head tracking (360° visual perception) | ✓ | | |
| | Virtual spatial navigation | ✓ | ✓ | |
| **Learning process** | Active learning (knowledge is triggered by interaction) | ✓ | ✓ | |
| | Passive learning (knowledge is fully represented) | | ✓ | ✓ |
| **Environment** | High immersion | ✓ | | |

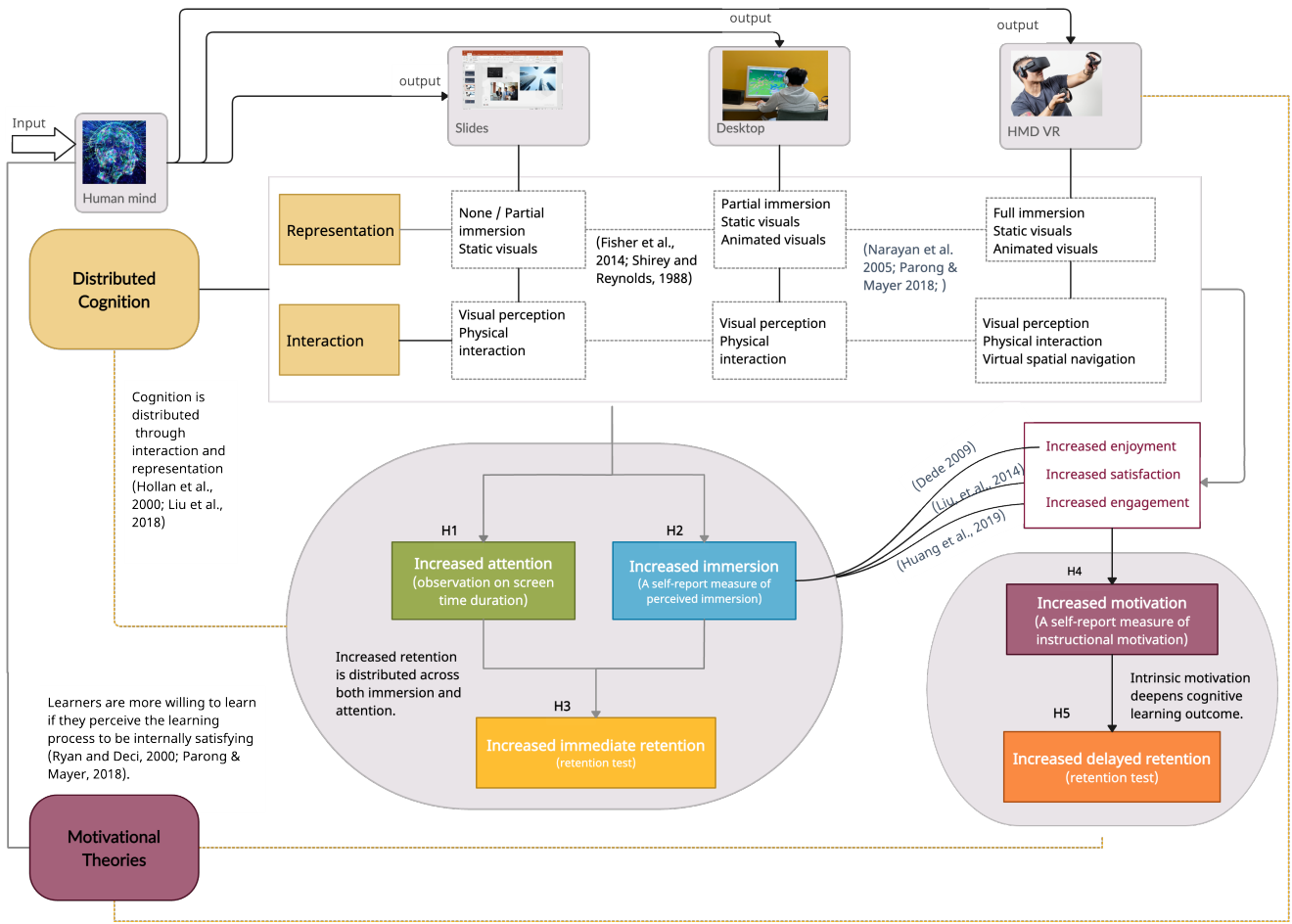

Figure 1: Conceptual Framework of the Research Design.

Although DCog does not account for whether one technological artifact is better than another, understanding how representation and interaction differ from each other can helps in understanding the learning process across different computer-mediated learning environments. Therefore, we analyzed the features of each instructional artifact that facilitated learning and categorized them as: representation, interaction, learning process, and environment, as described in Table 1. After comparing each feature of these technological artifacts, we identified three variables that account for representation and interaction differences among the varying instructional artifacts: learners' perceived immersion (accounting for learner's internal visual-spatial interaction and the external visual representation), screen time (accounting for learner's visual attention during cognitive processing), and learners' behavior and perceptions (accounting for physical interaction through gesture, movement, verbal and non-verbal behaviors). With this understanding of the mechanism of DCog, evaluating immersion, attention, learners' behavior, and perceptions allow us to understand how knowledge retention is constructed through representation and interaction in different learning environments.

Several studies indicated that sustained attention could be operationalized as screen time spent reading or watching instructional materials [25, 75, 100]. On this basis, a longer duration of screen time can predict a more motivated perception. Moreover, users' perceived immersion has been associated with a higher level of enjoyment, confidence, and positive attitudes [78, 83]. Meanwhile, *self-determination theory* posits that students would be more motivated to learn if their needs are met for autonomy and relatedness, referring to the fact that one would try hard in schoolwork if the learning material aligns with personal interests. Thus, we predicted that by using HMD VR, students would be more immersed and thus pay more visual attention during the lecture than using desktop and PowerPoint slides, which became our first two hypotheses ( **H1** and **H2**), as illustrated in Figure 1. Figure 1 also provides an overview of the conceptual framework that guides this research design and our theoretical conjectures.

### 3.2 Motivational Theories

Literature suggested that high levels of immersion [16, 78] and augmented visual perceptions [27] in VR are the main factors influencing users' learning engagement. According to both *interest theory* and *self-determination theory*, Learners are intrinsically motivated if their psychological needs are fulfilled. If one perceives the study activity and process to be internally satisfying, one would become self-determined and put more effort onto the learning materials. Furthermore, students are resilient in overcoming obstacles to learning [20, 51, 83]. The immersive nature of VR may draw learners' interest, and the interaction between them and the study content may sustain their engagement with the learning activities.

Levels of engagement can be measured through a user's emotional factors, such as attitude toward the content and satisfaction with the learning task. Ryan et al. [96] studied that the sense of immersion in participants' gaming experiences is associated with players' perceived competence and autonomy to predict their satisfaction needs. Huang et al. [41] sought to understand the role of motivation in virtual tourism and found that interactive activities relating to participants' psychological needs are essential to motivational dynamics in a virtual learning environment. By virtue of immersion and interactivity, carefully designed educational VR applications could contribute to an enhanced learning experience with the corresponding motivational perceptions to leverage cognitive gains. Therefore, students' instructional motivation can be predicted to improve using immersive VR compared to desktop applications or slides. Such increased motivation should continuously deepen students' cognitive processing even after VR experiences. By reviewing motivational theories and DCog together and using a customized immersive visualization as the learning material, the immediate retention of the concepts in the VR group was predicted to be higher than in the desktop group and slides group- this is our third hypothesis ( **H3**).

The perceived immersion and the visual attention spent on each lesson are expected to account for the difference in retention distribution. Thus, we proposed two additional hypotheses by reviewing motivational theories and DCog together and developing standard Food Microbiology lectures as immersive visualization content. We hypothesized that students using HMD VR would be more motivated to learn the subject ( **H4**) and achieve higher long-term retention after the learning activity ( **H5**) than those using desktop and slides.

## 4 CO-DESIGN

To further understand how VR facilitates learning from a student-centered perspective, we launched co-design approaches [14, 103] with relevant stakeholders to uncover pedagogical goals and design strategies for the Food Microbiology Laboratory lecture.

### 4.1 Design Stakeholders

Our co-design team is composed of two VR engineer developers, one Food Microbiology professor who has been teaching the class for decades, two students from the Food Microbiology department, and an education and HCI researcher.

### 4.2 The Process

The co-design sessions were conducted at the end of 2021 (5 months before the laboratory class began) to decide on appropriate instructional material for the VR lecture. After interviewing the instructor and several rounds of discussions, we chose "Lactic acid bacteria and sauerkraut Fermentation" as a learning material for the user study. This is a key curriculum module for many undergraduate Food Microbiology laboratory classes in the Unite States. The lecture covers the principles and process of fermentation, such as the roles of salt, cabbage, fermentation jar, and temperature in a kitchen. It also taught how different microorganisms' populations changed during fermentation. Usually, the lecture uses a classic fermentation curve chart to visualize a 27-30 day fermented process, and the lecture class last around half an hour. And then, in the later hands-on fermentation laboratory session, students will make their own sauerkraut fermentation and observe microbial populations based on the information retained from the lecture. Although the lecture is an indispensable part of the fermentation laboratory session, it is challenging for students to memorize complex terminologies, procedures, key data-set, and microbial properties within 20- 30 minutes.

The co-design team met several times for the instructional design and finalized the structure of the Fermentation VR lecture. Two virtual components were identified for learning objectives: The first component teaches the fermentation principles through a virtual kitchen setting so that students can virtually "enter" a kitchen to navigate each required ingredient for sauerkraut fermentation preparation. The research team modeled 3D objects in the kitchen environment and programmed interactive simulations. Students could explore each fermentation principle by reading text descriptions or listening to audio narration using a hand controller to interact virtually with relevant objects. The second component teaches microorganisms' properties, fermentation curve, and the pH value changes during the 30 days fermentation process. In this scenario, we created a virtual microbial world to show each microbe's actual population changes and color-encoded the virtual surrounding to show pH changes during fermentation. Students can use hand controllers to interact virtually with these microorganisms for an in-depth understanding of microbial properties, population changes, fermentation data, and pH values.

We presented our initial design idea with a set of storyboards. We then asked the instructor of the Food Microbiology Laboratory class and students to share their insights about the storyboards. We walked them through the workflow of the storyboard scenarios and invited them to identify design barriers and propose new interface strategies from a learner-centered perspective. Figure 2 shows an example storyboard during our design process. After several discussions within the research team, the final prototype is expected to comprise animated visuals, text descriptions, and audio narrations.

### 4.3 Two Prototypes Variations

The co-design approach guided us to create two variations of prototypes based on stakeholders' opinions. In particular, the variation is derived from whether students should be guided to learn or be given more freedom to explore study materials in VR [59, 112]. Because each type of pedagogy in interactive applications has certain benefits, two prototypes were created for pilot testing. Prototype 1 is an automated visualization application. The application is programmed to automatically display everyday fermentation information and has been set to visualize for 20 seconds per day. In this case, users can be immersed in the VR application to observe fermentation changes every 20 seconds. Prototype 2 is a user-driven visualization that allows users to have more interactivity choices during 30 days fermentation process. We added "previous" and "next" buttons to enable users to select a particular fermentation day for further exploration. An example of the two prototype interfaces is shown in Figure 3.

### 4.4 Pilot Study

We recruited 6 students with microbiology backgrounds to test the two prototypes on a within-subjects design basis. Among the participants, three were male, and three were female; 4 had no prior VR experiences, and 2 had light VR experiences. The pilot study lasted 1.5 hours for each participant and was conducted in a lab setting. Each session follows the procedures: 1). brief introduction and VR training session ( 10 minutes), testing prototype 1 ( 20 minutes), testing prototype 2 ( 20 minutes), filling out a questionnaire ( 20 minutes) and participating in a follow-up interview ( 20 minutes). The questionnaire consists of ten 7-point Likert scale questions adapted from the NASA-TLX [35] and System Usability Scale (SUS) [85] to evaluate the usability of two prototypes. During the follow-up interview, we also invited students to share their feedback about the prototypes and their suggestions regarding design updates.

### 4.5 Findings

A paired t-test revealed a significant difference ($t(6) = -3.34$, $p = 0.01$) between the usability of the system-automated prototype ($M = 5.08$; $SD = 0.53$) and the user-driven prototype ($M = 5.67$; $SD = 0.44$). This suggests that users preferred having more flexibility to make interactivity choices, with the user-driven prototype performing better than the system-automated prototype.

### 4.6 Design Strategies

The pilot study not only guided us in making decisions on the existing prototypes but also provided valuable feedback on our design choice for the successive iterations regarding improving the student's learning experiences. We analyzed interviews with participants and identified the following design strategies from the pilot participants' perspectives. We then applied these strategies to our application design update.

- *Balance the boundary between learning and play.* While all participants mentioned they prefer more flexibility to interact with the VR application, they also emphasized the importance of using moderate interaction with study materials to avoid potential confusion that might distract attention.
- *Support students' learning preferences to assist their information processing.* Prior research reveals that visual, auditory, and kin-aesthetic are the three most common ways students absorb information from lectures [32, 43]. Our interview with pilot study participants indicates their preference for both auditory and visual learning.
- *Ensuring text explicitly in immersive learning.* All participants mentioned text is still the main perceptual channel for their cognitive processing in VR. Therefore, it is crucial to ensure all text displays are clearly shown in students' view when using hand controllers to interact with VR content.

## 5 FERMENTATION VR

### 5.1 System Overview

Following these design strategies, we made revisions and implemented *Fermentation VR*, an immersive visualization application for teaching Sauerkarut Fermentation in the Food Microbiology laboratory class. *Fermentation VR* presents several features that immerse students in a food microbial-related virtual space to explore fermentation ingredients, environments, microorganisms, procedures, and processes.

As is shown ins Figure 4, *Fermentation VR* consists of three scenes: (1) an introduction to the VR lecture content, (2) a kitchen scene that allows students to explore different ingredients for fermentation preparation and principles, and (3) a microbial-view scene from the view of a fermentation jar to visualize the fermentation process.

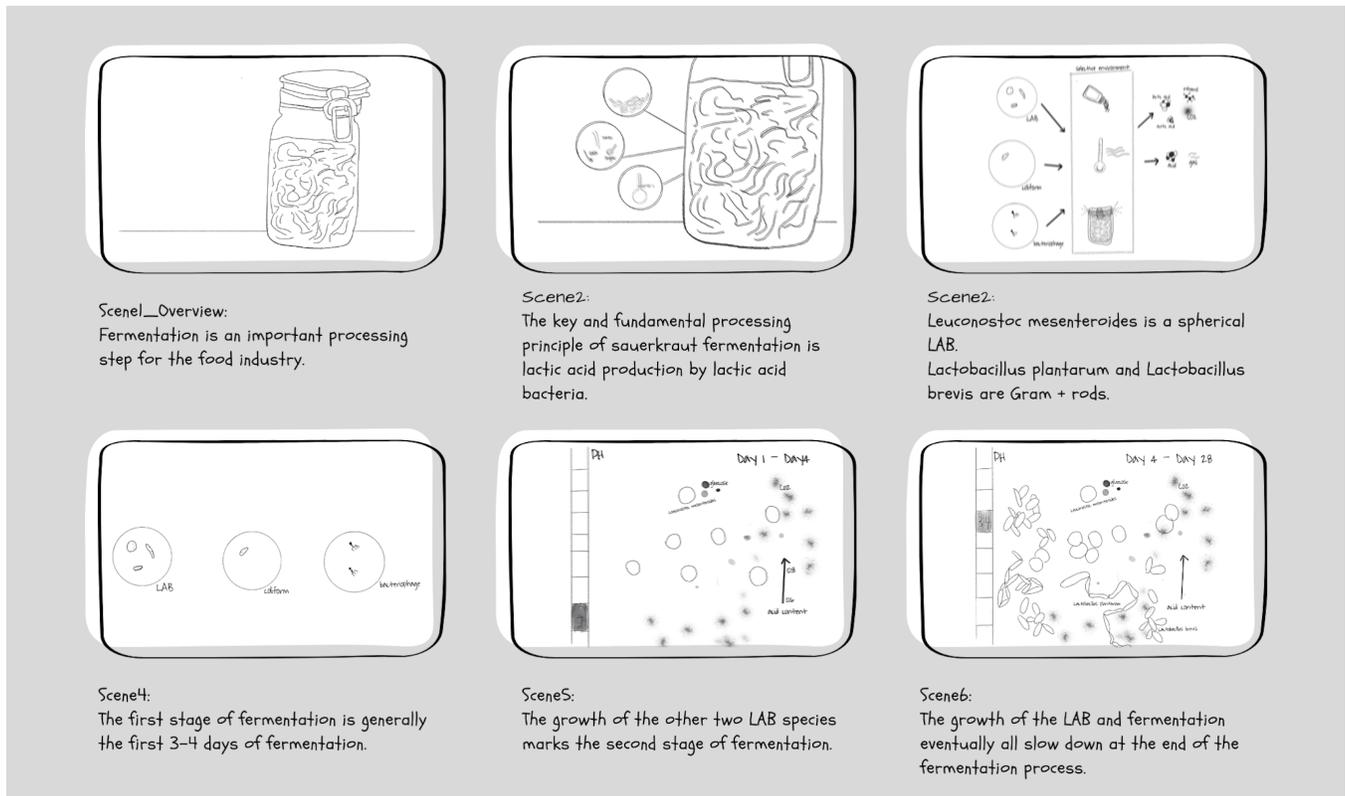

Figure 2: An example of storyboards developed during the co-design process.

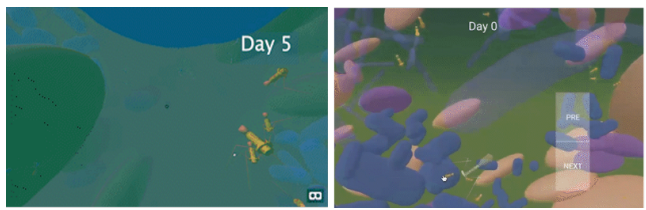

Figure 3: System automated prototype (left) versus user-driven prototype (right).

## 5.2 Implementation of VR, Desktop, and Slides

After the VR lesson was developed for immersive VR deployment, the same learning information was adapted into a desktop version and PowerPoint slides. The desktop-based lecture uses the same animated and interactive visuals as the VR lecture as a web application. Students sit in front of a computer to interact with the desktop application using a mouse. The slides utilize screenshots from the VR lesson, and identical audio narration has been inserted into each slide. All three types of learning materials are self-paced. To ensure experimental validity, every lesson in each condition last approximately 20 minutes.

## 6 USER STUDY

### 6.1 Experimental Design

Our user study used a three-condition between-subject experimental design, where participants were randomly assigned to one of the three conditions:

- **HMD Group** Participants stood in the center of the tracking area, wore an Oculus Quest HMD with a display resolution of 1920×3664 pixels, and held one hand controller to experience the HMD VR-based fermentation lecture.
- **DKP Group** Participants were seated at a desk. A Mac laptop was positioned in front of the participant in the tracking area. Individuals wore headphones and used a mouse and keyboard to interact with the desktop application. The lecture was developed using an A-frame Web VR application [104]. The learning materials in the desk application are the same as the HMD VR application and the slides.
- **SLS Group** Participants were seated at a desk. A Mac laptop was positioned in front of the participant in the tracking area. Individuals wore headphones and used a mouse and keyboard to interact with the PowerPoint slides. The lecture includes 19 pages of slides, and the learning materials are the same as the desktop application and HMD VR.

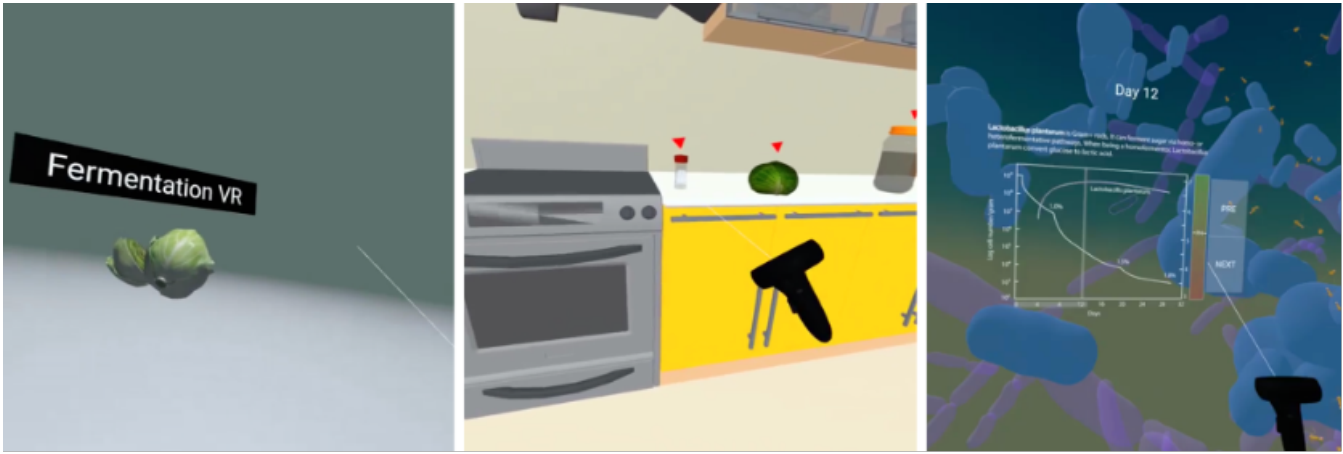

Figure 4: Interface scenarios. Left: Lecture introduction; Middle: Kitchen scene for fermentation preparation; Right: Microbial scene for fermentation processing.

## 6.2 Participants

The study took place at a large-sized USA university over five weeks of data collection during the Spring quarter of 2022. A total number of N = 49 undergraduate students enrolled in this Food Microbiology laboratory class participated in our research. They were randomly assigned to the HMD ( n = 17; 12% male; 22% female; age: $M = 23$; $SD = 4.58$), DKP (n = 16; 10% male; 22% female; age: $M = 22$; $SD = 1.05$), and SLS (n = 16; 12% male; 20% female; age: $M = 22$; $SD = 3.03$) conditions.

## 6.3 Overall Procedures

The experiment was a mandatory learning module from an introductory Food Microbiology Laboratory course for undergraduate microbiology majors. The course aims to train students in the laboratory methods used in the microbiological analysis of foods, which are critical mechanisms for studying microbial food processing, quality, and safety.

Upon approval by the Institutional Review Board, the primary researcher participated in the Food Microbiology Laboratory lecture class to debrief the project and recruit participants at the beginning of the quarter. After students provided consent to participate in the study, a pre-test was delivered during class time to understand students' prior knowledge. Then each student was scheduled individually with the primary researcher for their assigned self-directed fermentation lecture in a lab setting. A delayed post-test was delivered after one week of the user study during the class. It's worth noting that the study is voluntary basis and 7 students opted out of the research participants. An alternative lecture with the instructor is provided for these students to achieve the same learning objective. The experimental procedure is illustrated in Figure 5.

Before the experiment session, students provided their demographic information and general perspectives on educational VR in all conditions. During the experiment day, a training session was provided at the beginning for each participant to familiarize themselves with the functions of VR headsets and controllers. Then each participant was randomly assigned to each condition. In addition, participants were informed that they were free to take a break during the experimental lecture due to self-directed learning principles. After completing the lecture, participants filled out post-test 1 to measure their immediate retention of the learning gains. Then they answered a self-report questionnaire, including perceived immersion, instructional motivation, VR features, system usability, and learning outcomes. In the end, each participant was interviewed to reflect on their learning experiences and perceptions of HMD VR, desktop, and slides. Figure 6 illustrates how each condition was experimented with per session.

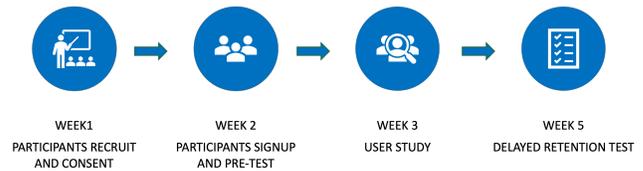

Figure 5: Project timeline description.

An identical delayed retention test (post-test 2) was given over one week of the experiments via Qualtrics survey software [88]. After students complete post-test 2, all students would complete a set of experimental fermentation tasks during the laboratory session. Students need to apply what they learned from the experimental lecture to their actual fermentation laboratory to ferment their own sauerkraut.

## 6.4 Experiment Measurements

*6.4.1 Background Information.* A survey was used to collect background information on students' demographics and prior VR experience. To reduce the confounding effects of the novelty of the VR technology, the research team offered two Google cardboard headsets to the participants at the beginning of the class to ensure there were no significant differences between students' prior user experiences of VR. In addition, students' previous knowledge of fermentation principles and processing was measured by a pre-test using the same content from post-test 1 and post-test 2.

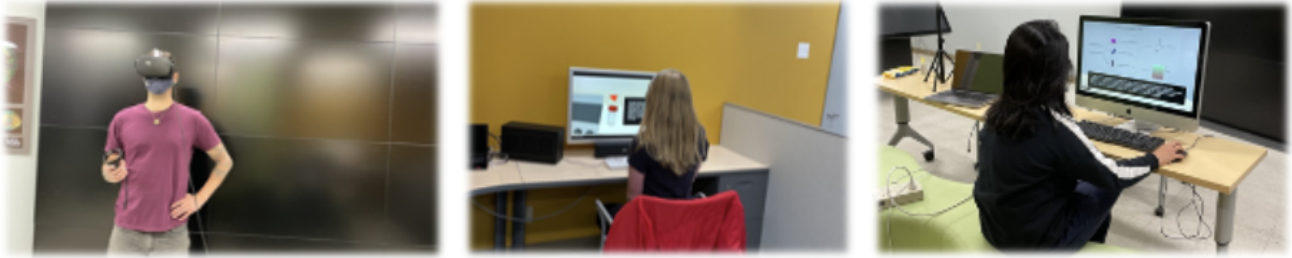

Figure 6: Three experimental conditions. From left to right: HMD VR, desktop application, and slides.

*6.4.2 Retention Test.* After the self-directed experimental lecture, students' knowledge of fermentation was measured by a post-test including eight items of multiple-choice questions. The class instructor and the educational researcher from the research team developed this retention test together based on the lecture content. Post-test 1 was given right after each experiment session to measure students' immediate retention. Post-test 2 was given after one week of the experimental lecture, and the score change between post-test 1 to post-test 2 was used to measure students' long-term retention.

*6.4.3 Visual Attention.* Screen time refers to the amount of time spent using a device with a screen. Previous literature [75] indicated that screen time is positively associated with an individual's self-perceived levels of visual attention. In a self-directed learning environment, assessing the total amount of time participants spent on the screen help to explore how visual representations distribute students' cognitive load and attention among different instructional artifacts.

*6.4.4 Perceived Immersion.* Students' perceived immersion was measured using an established immersion scale focusing on two constructs: absorptions and involvement. The scale is developed by Mütterlein [76] to understand the role of immersion in VR, and it consists of 6 likert rating questions.

*6.4.5 Instructional Motivation.* We used an adapted version of the Instructional Materials Motivation Survey (IMMS) developed by Kelly [47] to measure students' instructional motivation. This IMMS scale consists of nine statements concerning students' motivation for the instructional materials. The scale has been widely used in quantifying the extent of students' learning motivation by asking them to self-rate how positively and negatively they thought each statement aligns with their perceptions.

*6.4.6 Student's Perception of Instructional Artifacts.* Semi-structured interviews were used to deeply understand students' perceptions of using VR compared to desktop application and traditional slides lectures. Interview questions are mainly focused on asking students to reflect on their user and learning experiences with artifacts, the benefits and drawbacks of the artifacts, learning strategies used along with the learning process, and their suggestions for further improvements.

## 7 RESULTS

The following sections describe the analyses concerning the variables of retention test, self-reported immersion and instructional motivation, visual attention, and students' perceptions of using HMD VR, desktop, and slides.

### 7.1 Knowledge Retention

Table 2 shows the relative group-dependent means and SDs of the knowledge retention test scores before and after the experiments. Simple intragroup comparisons between the results from the pre-test and post-test1 revealed a significant increase (HMD: $t(16) = -3.21; p < 0.001$; DKP: $t(15) = -4.14; p < 0.001$; SLS: $t(15) = -3.62; p < 0.001$) of conceptual knowledge for all three groups, but no statistically significant difference (pretest: $F(2, 46) = 0.01; p = 0.99$; post-test1: $F(2, 46) = 0.54; p = 0.59$) among three groups in both pre-test and post-test1.

From post-test 1 to post-test 2, intragroup comparisons indicated a significant decrease (DKP: $t(15) = 2.50; p = 0.01$; SLS: $t(15) = 3.42; p = 0.001$) of conceptual knowledge for DKP and SLS groups, but no significant difference in HMD group ($t(16) = 0.18; p = 0.43$). Although no statistically significant difference ($F(2, 46) = 2.31; p = 0.11$) was identified among the three groups in post-test2, the retention level changes from post-test1 to post-test2 ($F(2, 46) = 3.78; p = 0.03$) showed that HMD VR could sustain learner' long-term retention better than desktop and slides.

We then conducted Tukey's HSD Test and Two-Way Repeated Measures ANOVA for further exploration. The post-hoc testing revealed that the mean value of retention change was significantly different between HMD and SLS groups ($p = 0.04$, 95% CI = [0.02, 2.07]). A two-way ANOVA with repeated measures revealed that there was not a statistically significant interaction between the effects of time (post-test1 and post-test2) and instructional artifacts ($F(2, 97) = 2.23, p = 0.11$). Simple main effects analysis showed that time has a statistically significant effect on retention change ($p = 0.003$). Simple main effects analysis showed that instructional artifacts don't have a statistically significant effect on retention change ($p = 0.60$).

### 7.2 Visual Attention

Observed screen time was used to measure participants' visual attention during self-directed learning in each experiment. A one-way ANOVA analysis indicated a significant difference in screen time ($F(2, 46) = 3.46; p = 0.04$) among the three groups (HMD:

Table 2: Mean comparison on pretest, post-test1, post-test2, visual attention, perceived immersion, instructional motivation, and score change over time (one week).

| Variables | Conditions | M | SD | F | $p$ | $\eta^2$ |
|---|---|---|---|---|---|---|
| **Pre-test Score** (7 is the full points) | HMD VR | 2.24 | 1.16 | 0.01 | 0.99 | 0 |
| | Desktop | 2.25 | 1.27 | | | |
| | Slides | 2.19 | 1.29 | | | |
| **Visual attention** (seconds) | HMD VR | 806 | 121 | 3.46 | 0.04* | 0.14 |
| | Desktop | 696 | 127 | | | |
| | Slides | 708 | 127 | | | |
| **Perceived Immersion** (7 is the full points) | HMD VR | 4.81 | 0.49 | 4.84 | 0.01* | 0.18 |
| | Desktop | 4.18 | 1.13 | | | |
| | Slides | 3.66 | 1.24 | | | |
| **Post-test 1 Score** (7 is the full points) | HMD VR | 3.34 | 0.74 | 0.54 | 0.59 | 0.03 |
| | Desktop | 3.75 | 1.08 | | | |
| | Slides | 3.59 | 1.02 | | | |
| **Instructional Motivation** (7 is the full points) | HMD VR | 5.36 | 0.68 | 4 | 0.03* | 0.15 |
| | Desktop | 5.19 | 0.53 | | | |
| | Slides | 4.73 | 0.62 | | | |
| **Post-test 2 Score** (7 is the full points) | HMD VR | 3.29 | 0.79 | 2.31 | 0.11 | 0.05 |
| | Desktop | 2.75 | 0.99 | | | |
| | Slides | 2.5 | 1.14 | | | |
| **Retention Score Change** (negative number refers retention decrease over 1 week) | HMD VR | -0.04 | 0.78 | 3.78 | 0.03* | 0.15 |
| | Desktop | -1.00 | 1.36 | | | |
| | Slides | -1.09 | 1.18 | | | |

$M = 806$, $SD = 121$; DKP: $M = 696$, $SD = 127$; SLS: $M = 708$, $SD = 127$). In the Tukey HSD post-hoc test, HMD participants exhibited longer screen time duration than DKP participants ($p = 0.04$, 95% CI = [-1.74, 222.48]) significantly. Although post-hoc test revealed no significant difference in visual attention between HMD and SLS groups, an independent t-test indicated a statistical difference ($t(31) = 2.41$, $p = 0.01$) between the two groups, suggesting that HMD participants spend more screen time than SLS participants.

### 7.3 Immersion and Instructional Motivation

To reveal significant differences in perceived immersion and instructional motivation, a one-way ANOVA analysis was applied to the average score. There was a significant group difference in perceived immersion ($F(2, 46) = 4.84$; $p = 0.01$) among three groups (HMD: $M = 4.81$, $SD = 0.49$; DKP: $M = 4.18$, $SD = 1.13$; SLS: $M = 3.66$, $SD = 1.24$). In the Tukey HSD post-hoc test, HMD participants revealed higher perceived immersion than DKP and SLS participants (HMD-DKP: $p < 0.001$; 95% CI = [3.06, 5.02]), HMD-SLS: $p < .001$; 95% C.I. = [2.61, 4.61]). A significant group difference in instructional motivation ($F(2, 46) = 4.00$; $p = 0.03$) was found among three groups (HMD: $M = 5.36$, $SD = 0.68$; DKP: $M = 5.19$, $SD = 0.53$; SLS: $M = 4.73$, $SD = 0.62$). The post-hoc testing also revealed that the mean value of HMD participants' instructional motivation was significantly different between HMD and SLS groups ($p = 0.02$, 95% CI = [0.07, 1.18]).

### 7.4 Correlation among Score Change, Immersion, Motivation, and Visual Attention

To explore the relationship between immersion, motivation, visual attention (screen time), and score change, the association among all variables was examined using Pearson's correlation analyses. Several significant correlations were found in this study.

First, significant correlations were found between score change and motivation, score change and immersion, immersion and motivation, immersion and screen time, motivation and screen time. There is a significant small-medium positive relationship between score change and immersion ($r(47) = 0.36$, $p = 0.01$); as well as score change and motivation ($r(47) = 0.30$, $p = 0.04$). Both immersion and motivation, and immersion and screen time were found to be strongly correlated: immersion and motivation ($r(47) = 0.70$, $p < 0.001$), immersion and screen time ($r(47) = 0.55$, $p < 0.001$). Motivation and screen time were found to be medium positively correlated with each other ($r(47) = 0.35$, $p = 0.02$). These results of the Pearson correlation indicated that students with higher immersion and motivation tend to better memorize the learning materials during the lecture experiment.

### 7.5 Students' Retention Score Change Over Time

We then analyzed the change in students' retention scores from pre-test to post-test 1 and post-test 2 per question (Q1 to Q7), as is shown in Figure 7. The figure shows no obvious unique patterns reflecting students' retention change from the three tests. However, it is worth noting that most questions (Q1, Q2, Q5, and Q7) have a higher retention rate from post-test 1 to post-test 2 in the HMD group compared to the other two groups. Particularly in Q5, the HMD groups' accuracy rate significantly increased from post-test1 to post-test 2, the SLS group's accuracy rate dropped from post-test 1 to post-test 2, and the DKP group's accuracy remained relatively slow increase from post-test 1 to post-test 2. Another interesting finding is that HMD students scored the highest points (0.96) in post-test1 to Q6, which refers to almost every student in the HMD group answering this question correctly, but scores dropped from post-test1 to post-test 2 in all three groups.

## 7.6 Students' Perceptions of Instructional Artifacts

In this section, we report our analysis of students' responses regarding their perceptions of HMD VR, desktop application, and slides during the interview.

*7.6.1 Thematic Analysis.* Under the guidance of established open coding methods [5], we performed a thematic analysis through an inductive process on all students' survey responses and interview transcripts. At first, the first author created a codebook of common themes concerning user experiences, benefits, challenges, and feedback as the first coder. Then another co-author coded all the transcripts and survey responses individually based on the initial code book. The research team discussed the emergent code through multiple coding passes until reaching a consensus on the themes reflecting students' perspectives. Lastly, we reached a strong agreement in inter-rater reliability with Cohen's Kappa $k > 0.6$.

*7.6.2 How do HMD VR, Desktop, and Slides Contribute to Learning?* In the post-questionnaire, we asked students how features of these instructional artifacts contribute to students' cognitive and affective processing. From the questionnaire responses, we found that students in the slides group feel the text is most valuable, but visuals are most interesting for their learning experiences. In contrast, the desktop group thought the text was most useful, and the virtual environment was most interesting for learning. However, it is interesting that the VR group thinks visual is most helpful for their learning than text, although they also feel the virtual environment is most interesting for their immersive learning experiences.

*7.6.3 Characteristics of Three Instructional Artifacts.* We transcribed the audio for interviews and then analyzed the transcripts using the MAXQDA qualitative analysis software [52]. In total, we collected 47 responses mentioning characteristics of HMD VR, 45 responses mentioning desktop applications, and 49 about slides. We started with an open coding process by segmenting the transcripts and applying descriptive codes [73]. Next, through an iterative analysis and discussion process, we looked for common features in which students acknowledged the artifacts' helpfulness and refined codes accordingly. After several rounds of discussion, the research team identified the characteristics of each instructional artifact from students' perspectives, as described in Table 3.

We found that 46% of comments shared that "engagement" is the main feature of HMD VR that helps with learning. In comparison, 24% of the comments highlighted "features/functionality" and 22% of comments mentioned "visualization" as a key characteristic of HMD VR for educational purposes. According to students' responses, staying engaged with learning materials in HMD VR is closely related to its enriched visual representations. According to several comments, such engagement makes students *"easy to focus on [study materials]"*. Features of HMD VR refer to the novel immersive experience in the virtual world, which is also closely related to the other two characteristics. Students' responses reveal that the novel immersive experiences and interactive visuals are the initial factors that trigger students' interests to continue exploring and thus foster a sense of sustained engagement through their learning experiences. For example, one student mentioned that: *"the process shows the material in pretty different and brand-new ways, the information pops up upon looking at it. I am not quite sure how to express the experiences, but I think in some way it is helping building up understandings of some materials"*.

Students' responses to what characteristics of desktop application help in learning are similar to their experiences with HMD VR. 37% of comments shared that "engagement" is the main characteristic of the desktop application for learning, while 33% of comments featured "interact with materials" and "features/functionality ." Desktop features mainly focus on the partial virtual experiences caused by "drag and rotate the mouse" and the personalized learning experience caused by "more user control to select information ." It is worth noting that "interact with materials" is more frequently mentioned as a beneficial characteristic of desktop application by students than "visualization." Although both desktop and VR implemented the same visual content, our interview results reveal that visual awareness is perceived more clearly when accompanied by a high level of perceived immersion.

When students shared their perceptions of slides, 50% of comments mentioned that the "ease of use" is the main characteristic of slides helps in learning; 44% of comments mentioned "navigation," and 31% of comments mentioned "familiarity." These findings reveal that usability is still essential for system designers and engineers to consider when designing novel educational technologies. Comments about navigation indicate the importance of efficiently accessing study materials in a self-directed learning environment. Moreover, "navigation" and "familiarity" also align with Norman's design principle on Affordance, which depicts the link between things look and how they are used in a user-centered design perspective [80].

*7.6.4 Study Strategies.* The general study strategies of using HMD VR, desktop application, and slides for self-directed learning differed from interviews with students and observed learning behavior during experiments. For example, Slides participants were observed to tend to skim all slides following a linear presentation sequence to get a quick overview and then jump to a particular slide for further investigation. On the contrary, we observed that desktop and HMD VR participants tend to navigate applications as a gaming experience through a lens of experiential learning to deepen their cognitive gains. For example, desktop and HMD VR participants used mouses and controllers to interact with virtual elements frequently regardless of whether the virtual object was related to the study materials. These differences are closely related to the characteristics of each instructional artifact and can be used to inform design principles for immersive learning.

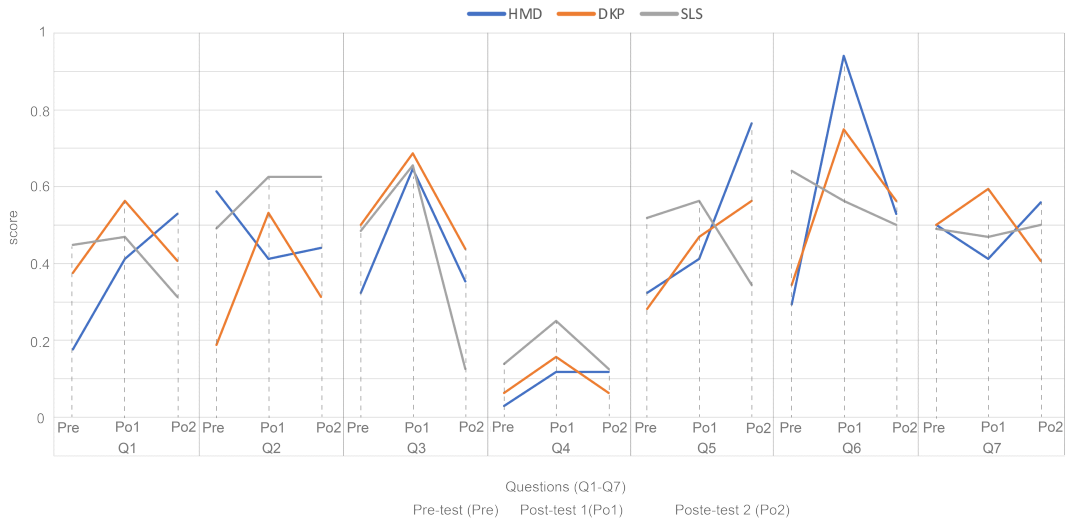

**Figure 7: Comparison of students' retention score in pretest, post-test 1, and post-test 2 per question.**

While the educational affordances of VR are well established, it is important to consider learners' classroom routine learning strategies. For example, during our study, when asking students if they would like to choose HMD VR instead of traditional slides for laboratory lectures, most students responded with *"no"* and *"I am not sure."* In addition, while students acknowledged VR's affordances in supporting engagement, many of them expressed that they are still accustomed to PowerPoint slides because this is '*"what I usually do for exam preparation."* These differences are closely related to the characteristics of each instructional artifact and can be used to inform design principles for immersive learning.

## 8 DISCUSSION, DESIGN IMPLICATIONS, AND LIMITATIONS

In this section, we reflect on our findings and provide insights into whether or not using VR to design effective learning experiences in laboratory classrooms in higher education.

### 8.1 Reflection on the Findings

Our results suggest the importance of considering the environmental challenges and design elements of using HMD VR for learning purposes. Particularly in our research, considering we measured participants' visual attention by screen time, students would be forced in attentive to learning content if they wore HMD. To address this bias, we chose the self-directed learning method to allow all participants to have the freedom to take a rest (e.g., taking off HMD and being away from the computer) during the experimental session and calculated the total screen time each participant spent. Hence, our result implies that more visual attention is paid due to participants' inherent enjoyment and perceived immersion, supporting our **H1** and **H2**. Our experiments indicate all three groups scored similarly on their immediate retention tests. Though this result seems discordant with **H3**, such non-homogeneous results don't reveal a false theoretical conjecture. One possible explanation is that note-taking was not allowed during the experiments.

Note-taking is commonly used in formal classroom lectures and aids short-term memory [93]. Therefore, when students can not use this familiar tool during experiments, they may experience anxiety, which could impact their ability to recall information quickly. We also found HMD VR participants' average instructional motivation was the highest among the three groups, theoretically supporting **H4**. Though no statistical difference in instructional motivation was identified between HMD VR and desktop groups, this might be attributed to the adverse effects of motion sickness [18, 56] and the discomfort of wearing HMD [6, 8, 57]. A few HMD VR participants reported discomfort during the experiment, and such an unpleasant perception may hinder students' desire to learn.

In addition, our study adds value to the theoretical literature. We identified that students increased curiosity and immersion can raise their intrinsic motivation to learn the subject. This expands Makransky's CAMIL model [63, 64] by suggesting that affective factors can facilitate conceptual knowledge acquisition. The result that HMD participants scored higher in long-term knowledge retention than other groups supports our **H5**. According to *interest theory*, one's situational interest in a study material can be translated into an individual interest in a long-term learning context [62]. Hence, our experiments demonstrate the integration of immersion, engagement, motivation, and attention as elements of interest that promote long-term retention. This finding, although slightly misaligned with the view that high immersion can harm cognitive processing [68, 84], highlights the importance of considering emotional factors for VR based learning. Our interview data indicate that high immersion environments may enhance students' visual awareness, which is consistent with recent studies [11, 65] showing a relationship between users' memory and kinematic performance with their visual mental images and spatial cognition in VR. Furthermore, our study provides support for the importance of theoretical guidance in VR research, as highlighted in Radianti et al.'s systematic review [89]. A solid theoretical foundation was crucial in all stages of the research process, including design, development, and evaluation.

Table 3: Summary of artifacts' characteristics helps in learning.

| Characteristics | Description | Example |
| --- | --- | --- |
| Ease of Use | Comments that describe the artifacts as being easy to use, simple, and/or convenient. | "The benefits of powerpoint slides are that they are easy to use and can be easily done in your own pace where you can go back and forth." |
| Features/Functionality | Comments related to the different features or functions that the artifact offered. e.g, immersion for VR, user control for desktop application, etc. | "The unconventional learning method which utilized VR, which is very uncommon. It allows the users to be fully involved in the virtual world that they are in with the materials and actions presented." |
| Access | Comments refer to the artifact being accessed or available for everyone to use. | "The benefits of the slides are easy access for students and faculty and they can be presented to a large audience." |
| Compatibility | Comments refer to how compatible the artifact was with another artifact. | "It(slides) is straightforwardly related with how we are tested (if we are tested in traditional ways). It is convenient as lecture note for exam review." |
| Navigation | Comments about the straightforward navigation or consistency of the artifact. | "They (slides) are easy to use and can be easily done in your own pace where you can go back and forth." |
| Familiarity | Comments refer to how familiar the artifact users were. | "Most everyone has a computer and know how to use it" |
| Visualization | Comments related to the visuals presented within the artifact. | "Can recall the information by moving my eyes to where it was in VR, I was able to see the microorganisms very clearly over time." |
| Interact with materials | Comments about the artifact facilitate interaction between the learner and the study material. | "Because the information pops up upon hovering with a mouse, it's easier to remember than with slides because it isn't monotonous." |
| Information processing | Comments about the use of the artifact to support cognitive processing of learning materials such as remembering, understanding, comprehension. | "The process show the material in pretty different and brand new ways. I am not quite sure how to express the experiences, but I think in someway it is helping building up understandings of some materials" |
| Engagement | Comments about the artifact being useful for promoting student engagement. | "I think VR is a lot more engaging which helped me keep my focus throughout the lesson. I was also more motivated to learn the material despite not needing to continue." |

Our research, along with recent related studies [17, 84, 113], indicates the importance of embodied activities and affective factors in instruction. For example, Innocenti et al. [17] found that spatial navigation in VR can reinforce students' interest in music genre learning and highlighted the necessity of evaluating the sustained impact of VR. In this regard, our work provides timely insights. However, Zhao et al. [113] found VR didn't outperform desktop display in Geography field trips. This may be attributed to inappropriate methods because the learning outcomes were evaluated via self-report questionnaires. In general, written formats using multiple-choice items is more reliable for VR lesson evaluation [60, 112, 113] as a self-reported questionnaire may lead to implicit bias. In terms of VR's impacts on students' long-term learning effects, our result is consistent with Kim et al. [49]'s finding in Medical education. In summary, reflecting on our results and in line with a few similar studies, VR could be used in higher education classrooms when appropriately designed [71]. On the one hand, designing authentic learning in VR that mimics how students behave in a real classroom is a key consideration. Although engagement is an essential characteristic of technological artifacts such as VR and desktop applications, our qualitative results indicate that students primarily seek instructional affordances from their daily activities and real-world experiences. Therefore, when designing effective VR learning experiences, future research should carefully reference conventional classroom practices. On the other hand, the theoretical framework proposed in our study has been validated and can be adapted to explore relationships between other emotional factors and learning-related behaviors when integrating technology into classrooms. Conducting such theory-grounded studies is imperative to identify appropriate design elements (e.g., academic domains, participants' individual differences, testing formats, etc.) for generalizable VR in higher education [18, 89].

## 8.2 Design Implications

*8.2.1 Visual Awareness through Embodiment is the Key for Learning Processing in VR.* Schmidt-Weigand et al. [97] found that the distribution of visual attention in digital learning is extensively guided by written text. Our study confirms this finding and emphasizes the importance of visual and text content for guiding learners' attention in a VR learning environment. When participants experience desktop display, visual awareness is less inter-weaved with text information processing because these visual changes are triggered by relatively abstract navigation interfaces, such as mouse-clicking and button-pressing [24, 39, 113]. However, in the fully immersive VR space, bodily actions such as walking and head rotation allow participants to perceive a higher level of visual awareness. These body-specific cues enable participants to foster a sense of embodiment, allowing them to access the virtual kitchen for sauerkraut fermentation as if they were in a fermentation jar experiencing microbial procedures from a kinesthetic perspective.

*8.2.2 Integrating Note-taking Feature into VR Lectures can Strengthen Students' Conceptual Understanding.* As mentioned earlier, note-taking was identified as a primary expectation that students hope to integrate into formal VR lectures during our interviews. For example, one student shared: *"I am used to taking notes when reading a slide, and it really helps me to reconstruct my understanding... but VR seems incapable of doing this because it is fully immersive."* This reminded us to reflect that the motivation for using slides in educational settings is not only a singular teaching method but also supplemental study materials along with writing notes. Furthermore, as an important approach to augmenting an individual's memory, note-taking allows students to categorize information as variables and store it in textual format for future reference. Integrating note-taking features into VR applications enables students to engage in deep information processing and achieve their full potential. In addition to facilitating memory recall, recent research has also shown the positive impacts of note-taking applications on academic success, in encouraging individual learning responsibility [114], increasing participants' social interaction quality [22, 30], and assimilating large information [79, 92].

*8.2.3 Non-linear Navigation should be Developed to Facilitate Students' Learning Preferences in VR .* Self-directed learning with technology is based on learners' preferred personal strategies. Therefore, it is necessary to take the initiative to consider learners' needs. Due to the complexity of the VR learning interface, knowledge cannot be presented in whole at once. Therefore, in VR, information is usually triggered by interactions between the learner and the explicit instructional content. This might lead to learners' negative learning experiences if they were unfamiliar with the interaction techniques during their immersive activity. Previous research reveals that when using PowerPoint slides, directly navigating to a particular slide provides rich leaner-content interaction [111] and helps students form concept relationships [6, 91]. Therefore, to ensure that the design principles do not impede the expected instructional effectiveness, it is crucial to consider navigation style while designing a VR lecture. Utilizing insights derived from PowerPoint affordances (e.g., ease of use, direct navigation, and user-friendly control buttons) can result in effective interaction design between study materials and learners.

*8.2.4 VR is not the Best Choice for all Learning Problems and Needs to Account for Realistic Classroom Situations.* Though VR has shown many benefits in teaching in higher education [57, 71, 74], our study identified that HMD VR doesn't facilitate students' better immediate retention than slides and desktop applications. From our qualitative analysis, we found that cost and technological access are primary concerns students have regarding introducing VR into formal classrooms, which has also been identified from recent work [33, 45]. These findings suggest that VR may not be the best choice for all learning problems. For example, the instructional value of VR is demonstrated in facilitating reflective thinking among students in chemistry [19], but not in the transfer of knowledge in history classrooms [84]. Moreover, to align with what we discussed in the Results section, although we identified that HMD VR helped students better maintain their memorization of fermentation concepts compared to slides and desktop display, only a few students are willing to use VR-based lectures in the formal Food Microbiology laboratory class. This suggests that the integration of VR technology in a real classroom setting is still in its early stages and requires further consideration for equity from all stakeholders involved from various stakeholders [45]. Therefore, we recommend whether or not using VR for higher education classrooms should be carefully tailored based on the learning objectives, instructors' and institutional support, students' attitudes toward technology, etc.

## 8.3 Limitations

Although this study has attempted to consider all aspects of the learning process and outcomes for using VR, desktop, and slides as detailed as possible, there are still limitations that are beyond the scope of the research. For example, we were not able to run the study with all students enrolled in the laboratory due to a voluntary basis which might reduce the statistical power in short-term and long-term learning effects. Secondly, although our study was conducted in a fundamental Food Microbiology laboratory class, we did not examine the transfer effects of students' lecture learning on their relevant laboratory experimental performances. A follow-up study is needed to further explore students' performance during the sauerkraut fermentation lab after immersive visualization. We intend to investigate this in our future work. Lastly, we only examined the interaction between learners and the technology (VR) from a distributed cognition perspective. How cognitive processes are distributed across learners in their daily activities and cultural practices within VR classrooms requires further exploration.

## 9 CONCLUSION AND FUTURE WORK

In our work, we proposed a theoretical framework to evaluate how immersive visualization facilitates knowledge retention grounded in the theory of distributed cognition and motivational theories. We then designed a VR lecture teaching the Fitness of Sauerkraut Fermentation by following co-design paradigms. Finally, we conducted a mixed-method study with Food Microbiology undergraduate students by randomly assigning them to three experimental groups: HMD VR, Desktop, and Slides. In particular, students' academic outcomes, learning experience data, and behavioral data were examined. Our findings suggest several design opportunities for HCI and education researchers, teachers, and VR application developers. Results presented in our study also provide insights for future improvements to advance the field.

We will continue to explore the impacts of different artifact-mediated lectures on students' subsequent hands-on lab performance. In doing so, we can develop an in-depth understanding of how declarative knowledge is transferred into procedural learning and explore the connection between long-term retention and transfer.

## ACKNOWLEDGMENTS

This research is sponsored in part by an Intel oneAPI Centers of Excellence grant. We thank Professor Lee Martin for his valuable feedback on the theory development and study design. We thank the late Tianchen Sun for his help with VR implementation. We also thank our participants for sharing their experiences during the user studies.

## REFERENCES
[1] Eytan Adar and Elsie Lee. 2020. Communicative visualizations as a learning problem. *IEEE Transactions on Visualization and Computer Graphics* 27, 2 (2020), 946–956. https://doi.org/10.1109/TVCG.2020.3030375
[2] Morana Alač and Edwin Hutchins. 2004. I see what you are saying: Action as cognition in fMRI brain mapping practice. *Journal of cognition and culture* 4, 3-4 (2004), 629–661.
[3] Ashlyn E Anderson, Louis B Justement, and Heather A Bruns. 2020. Using real-world examples of the COVID-19 pandemic to increase student confidence in their scientific literacy skills. *Biochemistry and Molecular Biology Education* 48, 6 (2020), 678–684. https://doi.org/10.1002/bmb.21474
[4] Sasha A Barab and Jonathan A Plucker. 2002. Smart people or smart contexts? Cognition, ability, and talent development in an age of situated approaches to knowing and learning. *Educational psychologist* 37, 3 (2002), 165–182. https://doi.org/10.1207/S15326985EP3703_3
[5] Michael J Belotto. 2018. Data analysis methods for qualitative research: Managing the challenges of coding, interrater reliability, and thematic analysis. *The Qualitative Report* 23, 11 (2018), 2622–2633.
[6] Andrew Brian Filmer and Ann Rosnida Md Deni. 2020. Powering Up Interactivity in the Use of PowerPoint within an Educational Context. In *2020 The 4th International Conference on Education and Multimedia Technology*. 107–111. https://doi.org/10.1145/3416797.3416845
[7] Graham Button. 1997. Cognition in the Wild, Edwin Hutchins. *Dordrecht* 6, 4 (Dec. 1997), 391–395.
[8] Fabio Buttussi and Luca Chittaro. 2017. Effects of different types of virtual reality display on presence and learning in a safety training scenario. *IEEE transactions on visualization and computer graphics* 24, 2 (2017), 1063–1076. https://doi.org/10.1109/TVCG.2017.2653117
[9] Jason A Chen, M Shane Tutwiler, Shari J Metcalf, Amy Kamarainen, Tina Grotzer, and Chris Dede. 2016. A multi-user virtual environment to support students' self-efficacy and interest in science: A latent growth model analysis. *Learning and Instruction* 41 (2016), 11–22. https://doi.org/10.1016/j.learninstruc.2015.09.007
[10] Alan Cheng, Lei Yang, and Erik Andersen. 2017. Teaching language and culture with a virtual reality game. In *Proceedings of the 2017 CHI conference on human factors in computing systems*. 541–549. https://doi.org/10.1145/3025453.3025857
[11] Xiangtong Chu, Xiao Xie, Shuainan Ye, Haolin Lu, Hongguang Xiao, Zeqing Yuan, Zhutian Chen, Hui Zhang, and Yingcai Wu. 2021. TIVEE: Visual exploration and explanation of badminton tactics in immersive visualizations. *IEEE Transactions on Visualization and Computer Graphics* 28, 1 (2021), 118–128. https://doi.org/10.1109/TVCG.2021.3114861
[12] Jennifer Clark. 2008. PowerPoint and pedagogy: Maintaining student interest in university lectures. *College teaching* 56, 1 (2008), 39–44.
[13] Amanda M Cottone and Susan Yoon. 2020. Improving the design of undergraduate biology courses toward the goal of retention: the case of real-world inquiry and active learning through metagenomics. *Journal of microbiology & biology education* 21, 1 (2020), 20. https://doi.org/10.1128/jmbe.v21i1.1965
[14] Giovanni De Michell and Rajesh K Gupta. 1997. Hardware/software co-design. *Proc. IEEE* 85, 3 (1997), 349–365.
[15] Edward L Deci and Richard M Ryan. 2008. Self-determination theory: A macrotheory of human motivation, development, and health. *Canadian Psychology/Psychologie canadienne* 49, 3 (Aug. 2008), 182–185. https://doi.org/10.1037/a0012801
[16] Chris Dede. 2009. Immersive interfaces for engagement and learning. *science* 323, 5910 (2009), 66–69. https://doi.org/10.1126/science.1167311
[17] Edoardo Degli Innocenti, Michele Geronazzo, Diego Vescovi, Rolf Nordahl, Stefania Serafin, Luca Andrea Ludovico, and Federico Avanzini. 2019. Mobile virtual reality for musical genre learning in primary education. *Computers & Education* 139 (2019), 102–117.
[18] Anna Flavia Di Natale, Claudia Repetto, Giuseppe Riva, and Daniela Villani. 2020. Immersive virtual reality in K-12 and higher education: A 10-year systematic review of empirical research. *British Journal of Educational Technology* 51, 6 (2020), 2006–2033. https://doi.org/10.1111/bjet.13030
[19] Cathi L Dunnagan, Devran A Dannenberg, Michael P Cuales, Arthur D Earnest, Richard M Gurnsey, and Maria T Gallardo-Williams. 2019. Production and evaluation of a realistic immersive virtual reality organic chemistry laboratory experience: infrared spectroscopy. https://doi.org/10.1021/acs.jchemed.9b00705
[20] Jacquelynne S Eccles, Allan Wigfield, et al. 2002. Motivational beliefs, values, and goals. *Annual review of psychology* 53, 1 (2002), 109–132.
[21] Mohamed El Beheiry, Sébastien Doutreligne, Clément Caporal, Cécilia Ostertag, Maxime Dahan, and Jean-Baptiste Masson. 2019. Virtual reality: beyond visualization. *Journal of molecular biology* 431, 7 (2019), 1315–1321.
[22] Marc Exposito, Vicky Zeamer, and Pattie Maes. 2017. Unobtrusive Note Taking: Enriching Digital Interpersonal Interactions Using Gestures. In *Companion of the 2017 ACM Conference on Computer Supported Cooperative Work and Social Computing*. 167–170. https://doi.org/10.1145/3022198.3026319
[23] Michel Fayol. 1994. From declarative and procedural knowledge to the management of declarative and procedural knowledge. *European journal of psychology of education* 9, 3 (1994), 179–190.
[24] Yan Feng. 2021. Do different types of VR influence pedestrian route choice behaviour? A comparison study of Desktop VR and HMD VR. In *Extended Abstracts of the 2021 CHI Conference on Human Factors in Computing Systems*. 1–7. https://doi.org/10.1145/3411763.3451516
[25] Anna V Fisher, Karrie E Godwin, and Howard Seltman. 2014. Visual environment, attention allocation, and learning in young children: When too much of a good thing may be bad. *Psychological science* 25, 7 (2014), 1362–1370. https://doi.org/10.1177/0956797614533801
[26] Xinyi Fu, Yaxin Zhu, Zhijing Xiao, Yingqing Xu, and Xiaojuan Ma. 2020. RestoreVR: generating embodied knowledge and situated experience of dunhuang mural conservation via interactive virtual reality. In *Proceedings of the 2020 CHI Conference on Human Factors in Computing Systems*. 1–13. https://doi.org/10.1145/3313831.3376673
[27] Hong Gao, Efe Bozkir, Lisa Hasenbein, Jens-Uwe Hahn, Richard Göllner, and Enkelejda Kasneci. 2021. Digital transformations of classrooms in virtual reality. In *Proceedings of the 2021 CHI Conference on Human Factors in Computing Systems*. 1–10. https://doi.org/10.1145/3411764.3445596
[28] Sarah Garcia, Ronald Kauer, Denis Laesker, Jason Nguyen, and Marvin Andujar. 2019. A virtual reality experience for learning languages. In *Extended Abstracts of the 2019 CHI Conference on Human Factors in Computing Systems*. 1–4. https://doi.org/10.1145/3290607.3313253
[29] J Georgiou, K Dimitropoulos, and A Manitsaris. 2007. A virtual reality laboratory for distance education in chemistry. *International Journal of Social Sciences* 2, 1 (2007), 34–41.
[30] Katy Ilonka Gero, Lydia Chilton, Chris Melancon, and Mike Cleron. 2022. Eliciting Gestures for Novel Note-taking Interactions. In *Designing Interactive Systems*

*Conference*. 966–975. https://doi.org/10.1145/3532106.3533480
[31] Jan Géryk. 2015. Using Visual Analytics Tool for Improving Data Comprehension. *International Educational Data Mining Society* (2015).
[32] Gilakjani. 2012. Visual, auditory, kinaesthetic learning styles and their impacts on English language teaching. *J. Stud. Educ.* (2012).
[33] Yu Han, Yu Miao, Jie Lu, Mei Guo, and Yi Xiao. 2022. Exploring Intervention Strategies for Distracted Students in VR Classrooms. In *CHI Conference on Human Factors in Computing Systems Extended Abstracts*. 1–7. https://doi.org/10.1145/3491101.3519627
[34] Jordan Harold, Irene Lorenzoni, Thomas F Shipley, and Kenny R Coventry. 2016. Cognitive and psychological science insights to improve climate change data visualization. *Nature Climate Change* 6, 12 (2016), 1080–1089.
[35] Sandra G Hart and Lowell E Staveland. 1988. Development of NASA-TLX (Task Load Index): Results of empirical and theoretical research. In *Advances in psychology*. Vol. 52. Elsevier, 139–183.
[36] Christopher Healey and James Enns. 2011. Attention and visual memory in visualization and computer graphics. *IEEE transactions on visualization and computer graphics* 18, 7 (2011), 1170–1188. https://doi.org/10.1109/TVCG.2011.127
[37] Paula Hodgson, Vivian WY Lee, Johnson Chan, Agnes Fong, Cindi SY Tang, Leo Chan, and Cathy Wong. 2019. Immersive virtual reality (IVR) in higher education: Development and implementation. In *Augmented reality and virtual reality*. Springer, 161–173. https://doi.org/10.1007/978-3-030-06246-0_12
[38] James Hollan, Edwin Hutchins, and David Kirsh. 2000. Distributed cognition: toward a new foundation for human-computer interaction research. *ACM Transactions on Computer-Human Interaction (TOCHI)* 7, 2 (2000), 174–196. https://doi.org/10.1145/353485.353487
[39] Jan Hombeck, Monique Meuschke, Lennert Zyla, André-Joel Heuser, Justus Toader, Felix Popp, Christiane J Bruns, Christian Hansen, Rabi R Datta, and Kai Lawonn. 2022. Evaluating Perceptual Tasks for Medicine: A Comparative User Study Between a Virtual Reality and a Desktop Application. In *2022 IEEE Conference on Virtual Reality and 3D User Interfaces (VR)*. IEEE, 514–523. https://doi.org/DOI:10.1109/VR51125.2022.00071
[40] Tien-Chi Huang, Chia-Chen Chen, and Yu-Wen Chou. 2016. Animating eco-education: To see, feel, and discover in an augmented reality-based experiential learning environment. *Computers & Education* 96 (2016), 72–82. https://doi.org/10.1016/j.compedu.2016.02.008
[41] Yu-Chih Huang, Sheila J Backman, Kenneth F Backman, Francis A McGuire, and DeWayne Moore. 2019. An investigation of motivation and experience in virtual learning environments: A self-determination theory. *Education and Information Technologies* 24, 1 (2019), 591–611.
[42] Edwin Hutchins. 1991. The social organization of distributed cognition. American Psychological Association.
[43] Radhwan Hussein Ibrahim and Dhia-Alrahman Hussein. 2016. Assessment of visual, auditory, and kinesthetic learning style among undergraduate nursing students. *Int J Adv Nurs Stud* 5, 1 (2016), 1–4.
[44] Crescent Jicol, Chun Hin Wan, Benjamin Doling, Caitlin H Illingworth, Jinha Yoon, Charlotte Headey, Christof Lutteroth, Michael J Proulx, Karin Petrini, and Eamonn O'Neill. 2021. Effects of emotion and agency on presence in virtual reality. In *Proceedings of the 2021 CHI Conference on Human Factors in Computing Systems*. 1–13. https://doi.org/10.1145/3411764.3445588
[45] Qiao Jin, Yu Liu, Svetlana Yarosh, Bo Han, and Feng Qian. 2022. How Will VR Enter University Classrooms? Multi-stakeholders Investigation of VR in Higher Education. In *CHI Conference on Human Factors in Computing Systems*. 1–17. https://doi.org/10.1145/3491102.3517542
[46] Elizabeth Johnston, Gerald Olivas, Patricia Steele, Cassandra Smith, and Liston Bailey. 2018. Exploring pedagogical foundations of existing virtual reality educational applications: A content analysis study. *Journal of Educational Technology Systems* 46, 4 (2018), 414–439. https://doi.org/10.1177/0047239517745560
[47] John M Keller. 1987. IMMS: Instructional materials motivation survey. *Florida State University* (1987).
[48] Mina Khan, Fernando Trujano, and Pattie Maes. 2018. Mathland: Constructionist mathematical learning in the real world using immersive mixed reality. In *International Conference on Immersive Learning*. Springer, 133–147. https://doi.org/10.1007/978-3-319-93596-6_9
[49] Seunghyun Kim, Ryoun Heo, Yeonji Chung, Jung Min Kim, Michelle P Kwon, Sung Chul Seo, Gil-Hong Park, and Meyoung-Kon Kim. 2019. Virtual reality visualization model (VRVM) of the tricarboxylic acid (TCA) cycle of carbohydrate metabolism for medical biochemistry education. *Journal of Science Education and Technology* 28, 6 (2019), 602–612. https://doi.org/10.1007/s10956-019-09790-y
[50] Pascal Knierim, Valentin Schwind, Anna Maria Feit, Florian Nieuwenhuizen, and Niels Henze. 2018. Physical keyboards in virtual reality: Analysis of typing performance and effects of avatar hands. In *Proceedings of the 2018 CHI Conference on Human Factors in Computing Systems*. 1–9. https://doi.org/10.1145/3173574.3173919
[51] Jordan Koulouris, Zoe Jeffery, James Best, Eamonn O'neill, and Christof Lutteroth. 2020. Me vs. Super (wo) man: Effects of Customization and Identification in a VR Exergame. In *Proceedings of the 2020 CHI Conference on Human Factors in Computing Systems*. 1–17. https://doi.org/10.1145/3313831.3376661
[52] Udo Kuckartz and Stefan Rädiker. 2019. *Analyzing Qualitative Data with MAXQDA: Text, Audio, and Video*. Springer, Cham.
[53] Oh-Hyun Kwon, Chris Muelder, Kyungwon Lee, and Kwan-Liu Ma. 2016. A study of layout, rendering, and interaction methods for immersive graph visualization. *IEEE transactions on visualization and computer graphics* 22, 7 (2016), 1802–1815. https://doi.org/10.1109/TVCG.2016.2520921
[54] Joseph J LaViola Jr, Ernst Kruijff, Ryan P McMahan, Doug Bowman, and Ivan P Poupyrev. 2017. *3D user interfaces: theory and practice*. Addison-Wesley Professional.
[55] Elinda Ai-Lim Lee, Kok Wai Wong, and Chun Che Fung. 2010. How does desktop virtual reality enhance learning outcomes? A structural equation modeling approach. *Computers & Education* 55, 4 (2010), 1424–1442. https://doi.org/10.1016/j.compedu.2010.06.006
[56] Jeonghaeng Lee, Woojae Kim, Jinwoo Kim, and Sanghoon Lee. 2021. A Study on Virtual Reality Sickness and Visual Attention. In *2021 Asia-Pacific Signal and Information Processing Association Annual Summit and Conference (APSIPA ASC)*. IEEE, 1465–1469.
[57] Ruixue Liu, Lei Wang, Jing Lei, Qiu Wang, and Youqun Ren. 2020. Effects of an immersive virtual reality-based classroom on students' learning performance in science lessons. *British Journal of Educational Technology* 51, 6 (2020), 2034–2049. https://doi.org/10.1111/bjet.13028
[58] Zhicheng Liu, Nancy Nersessian, and John Stasko. 2008. Distributed cognition as a theoretical framework for information visualization. *IEEE transactions on visualization and computer graphics* 14, 6 (2008), 1173–1180. https://doi.org/10.1109/TVCG.2008.121
[59] Andreas Löcken, Carmen Golling, and Andreas Riener. 2019. How should automated vehicles interact with pedestrians? A comparative analysis of interaction concepts in virtual reality. In *Proceedings of the 11th international conference on automotive user interfaces and interactive vehicular applications*. 262–274. https://doi.org/10.1145/3342197.3344544
[60] Michelle Lui, Rhonda McEwen, and Martha Mullally. 2020. Immersive virtual reality for supporting complex scientific knowledge: Augmenting our understanding with physiological monitoring. *British Journal of Educational Technology* 51, 6 (2020), 2181–2199. https://doi.org/10.1111/bjet.13022
[61] Kwan-Liu Ma, I Liao, J Frazier, H Hauser, and H-N Kostis. 2012. Scientific Storytelling Using Visualization. *IEEE Computer Graphics and Applications* 32, 1 (2012), 12–19.
[62] Ulrike IE Magner, Rolf Schwonke, Vincent Aleven, Octav Popescu, and Alexander Renkl. 2014. Triggering situational interest by decorative illustrations both fosters and hinders learning in computer-based learning environments. *Learning and instruction* 29 (2014), 141–152. https://doi.org/10.1016/j.learninstruc.2012.07.002
[63] Guido Makransky and Gustav Bøg Petersen. 2019. Investigating the process of learning with desktop virtual reality: A structural equation modeling approach. *Computers & Education* 134 (2019), 15–30. https://doi.org/10.1016/j.compedu.2019.02.002
[64] Guido Makransky and Gustav B Petersen. 2021. The cognitive affective model of immersive learning (CAMIL): A theoretical research-based model of learning in immersive virtual reality. *Educational Psychology Review* 33, 3 (2021), 937–958. https://doi.org/10.1007/s10648-020-09586-2
[65] Katerina Mania, Dave Wooldridge, Matthew Coxon, and Andrew Robinson. 2006. The effect of visual and interaction fidelity on spatial cognition in immersive virtual environments. *IEEE transactions on visualization and computer graphics* 12, 3 (2006), 396–404. https://doi.org/10.1109/TVCG.2006.55
[66] David M Markowitz, Rob Laha, Brian P Perone, Roy D Pea, and Jeremy N Bailenson. 2018. Immersive virtual reality field trips facilitate learning about climate change. *Frontiers in psychology* 9 (2018), 2364. https://doi.org/10.3389/fpsyg.2018.02364
[67] Stefan Marks, Javier E Estevez, and Andy M Connor. 2014. Towards the Holodeck: fully immersive virtual reality visualisation of scientific and engineering data. In *Proceedings of the 29th International Conference on Image and Vision Computing New Zealand*. 42–47. https://doi.org/10.1145/2683405.2683424
[68] Richard E Mayer. 2005. Cognitive theory of multimedia learning. *The Cambridge handbook of multimedia learning* 41 (2005), 31–48.
[69] Joan M Mazur and Cindy H Lio. 2004. Learner articulation in an immersive visualization environment. In *CHI'04 Extended Abstracts on Human Factors in Computing Systems*. 1355–1358. https://doi.org/10.1145/985921.986063
[70] Huang Hsiu Mei and Liaw Shu Sheng. 2011. Applying situated learning in a virtual reality system to enhance learning motivation. *International journal of information and education technology* 1, 4 (2011), 298–302.
[71] Zahira Merchant, Ernest T Goetz, Lauren Cifuentes, Wendy Keeney-Kennicutt, and Trina J Davis. 2014. Effectiveness of virtual reality-based instruction on students' learning outcomes in K-12 and higher education: A meta-analysis. *Computers & Education* 70 (2014), 29–40. https://doi.org/10.1016/j.compedu.2013.07.033

[72] Susan Merkel and ASM Task Force on Curriculum Guidelines for Undergraduate Microbiology. 2012. The development of curricular guidelines for introductory microbiology that focus on understanding. *Journal of Microbiology & Biology Education* 13, 1 (2012), 32–38. https://doi.org/10.1128/jmbe.v13i1.363

[73] Sharan B Merriam and Elizabeth J Tisdell. 2015. *Qualitative Research: A Guide to Design and Implementation.* John Wiley & Sons.

[74] Oliver A Meyer, Magnus K Omdahl, and Guido Makransky. 2019. Investigating the effect of pre-training when learning through immersive virtual reality and video: A media and methods experiment. *Computers & Education* 140 (2019), 103603. https://doi.org/10.1016/j.compedu.2019.103603

[75] Ilaria Montagni, Elie Guichard, and Tobias Kurth. 2016. Association of screen time with self-perceived attention problems and hyperactivity levels in French students: a cross-sectional study. *BMJ open* 6, 2 (2016), e009089. https://doi.org/10.1136/bmjopen-2015-009089

[76] Joschka Mütterlein. 2018. The three pillars of virtual reality? Investigating the roles of immersion, presence, and interactivity. In *Proceedings of the 51st Hawaii international conference on system sciences.*

[77] Pranathi Mylavarapu, Adil Yalcin, Xan Gregg, and Niklas Elmqvist. 2019. Ranked-list visualization: A graphical perception study. In *Proceedings of the 2019 CHI Conference on Human Factors in Computing Systems.* 1–12. https://doi.org/10.1145/3290605.3300422

[78] Michael Narayan, Leo Waugh, Xiaoyu Zhang, Pradyut Bafna, and Doug Bowman. 2005. Quantifying the benefits of immersion for collaboration in virtual environments. In *Proceedings of the ACM symposium on Virtual reality software and technology.* 78–81.

[79] Cuong Nguyen and Feng Liu. 2016. Gaze-based notetaking for learning from lecture videos. In *Proceedings of the 2016 CHI Conference on Human Factors in Computing Systems.* 2093–2097. https://doi.org/10.1145/2858036.2858137

[80] Donald A Norman. 1988. The design of everything things. *Currency Doubleday, USA* (1988).

[81] Max M North, Joseph Sessum, and Alex Zakalev. 2004. Immersive visualization tool for pedagogical practices of computer science concepts: A pilot study. *Journal of Computing Sciences in Colleges* 19, 3 (2004), 207–215.

[82] Sebastian Oberdörfer, David Heidrich, and Marc Erich Latoschik. 2019. Usability of gamified knowledge learning in VR and desktop-3D. In *Proceedings of the 2019 CHI Conference on Human Factors in Computing Systems.* 1–13. https://doi.org/10.1145/3290605.3300405

[83] Jocelyn Parong and Richard E Mayer. 2018. Learning science in immersive virtual reality. *Journal of Educational Psychology* 110, 6 (2018), 785. https://doi.org/10.1037/edu0000241

[84] Jocelyn Parong and Richard E Mayer. 2021. Learning about history in immersive virtual reality: does immersion facilitate learning? *Educational Technology Research and Development* 69, 3 (2021), 1433–1451. https://doi.org/10.1007/s11423-021-09999-y

[85] S Camille Peres, Tri Pham, and Ronald Phillips. 2013. Validation of the system usability scale (SUS) SUS in the wild. In *Proceedings of the Human Factors and Ergonomics Society Annual Meeting*, Vol. 57. SAGE Publications Sage CA: Los Angeles, CA, 192–196. https://doi.org/10.1177/1541931213571043

[86] Johanna Pirker, Michael Holly, Isabel Lesjak, Johannes Kopf, and Christian Gütl. 2019. MaroonVR—An Interactive and Immersive Virtual Reality Physics Laboratory. In *Learning in a Digital World*. Springer, 213–238. https://doi.org/10.1007/978-981-13-8265-9_11

[87] Aaron M Pleitner, Susan R Hammons, Emily McKenzie, Young-Hee Cho, and Haley F Oliver. 2014. Introduction of molecular methods into a food microbiology curriculum. *Journal of Food Science Education* 13, 4 (2014), 68–76. https://doi.org/10.1111/1541-4329.12042

[88] Qualtrics, Provo, UT. 2020. Qualtrics | Analysis & Software. https://www.qualtrics.com. [Online; accessed 22-August-2022].

[89] Jaziar Radianti, Tim A Majchrzak, Jennifer Fromm, and Isabell Wohlgenannt. 2020. A systematic review of immersive virtual reality applications for higher education: Design elements, lessons learned, and research agenda. *Computers & Education* 147 (2020), 103778.

[90] Iulian Radu and Bertrand Schneider. 2019. What can we learn from augmented reality (AR)? Benefits and drawbacks of AR for inquiry-based learning of physics. In *Proceedings of the 2019 CHI conference on human factors in computing systems.* 1–12. https://doi.org/10.1145/3290605.3300774

[91] Stephen K Reed. 2006. Cognitive architectures for multimedia learning. *Educational psychologist* 41, 2 (2006), 87–98. https://doi.org/10.1207/s15326985ep4102_2

[92] Yolanda Jacobs Reimer, Erin Brimhall, Chen Cao, and Kevin O'Reilly. 2009. Empirical user studies inform the design of an e-notetaking and information assimilation system for students in higher education. *Computers & Education* 52, 4 (2009), 893–913. https://doi.org/10.1016/j.compedu.2008.12.013

[93] Yann Riche, Nathalie Henry Riche, Ken Hinckley, Sheri Panabaker, Sarah Fuelling, and Sarah Williams. 2017. As We May Ink?: Learning from Everyday Analog Pen Use to Improve Digital Ink Experiences. (2017), 3241–3253. https://doi.org/10.1145/3025453.3025716

[94] Maria Roussou. 2004. Learning by doing and learning through play: an exploration of interactivity in virtual environments for children. *Computers in Entertainment (CIE)* 2, 1 (2004), 10–10.

[95] Puripant Ruchikachorn and Klaus Mueller. 2015. Learning visualizations by analogy: Promoting visual literacy through visualization morphing. *IEEE transactions on visualization and computer graphics* 21, 9 (2015), 1028–1044. https://doi.org/10.1109/TVCG.2015.2413786

[96] Richard M Ryan, C Scott Rigby, and Andrew Przybylski. 2006. The motivational pull of video games: A self-determination theory approach. *Motivation and emotion* 30, 4 (2006), 344–360. https://doi.org/10.1007/s11031-006-9051-8

[97] Florian Schmidt-Weigand, Alfred Kohnert, and Ulrich Glowalla. 2010. A closer look at split visual attention in system- and self-paced instruction in multimedia learning. *Learning and instruction* 20, 2 (2010), 100–110.

[98] Christian Schott and Stephen Marshall. 2018. Virtual reality and situated experiential education: A conceptualization and exploratory trial. *Journal of computer assisted learning* 34, 6 (2018), 843–852. https://doi.org/10.1111/jcal.12293

[99] Jinsil Hwaryoung Seo, Brian Michael Smith, Margaret E Cook, Erica R Malone, Michelle Pine, Steven Leal, Zhikun Bai, and Jinkyo Suh. 2017. Anatomy builder VR: Embodied VR anatomy learning program to promote constructionist learning. In *Proceedings of the 2017 CHI Conference Extended Abstracts on Human Factors in Computing Systems.* 2070–2075. https://doi.org/10.1145/3027063.3053148

[100] Larry L Shirey and Ralph E Reynolds. 1988. Effect of interest on attention and learning. *Journal of Educational Psychology* 80, 2 (1988), 159.

[101] Yu Shu, Yen-Zhang Huang, Shu-Hsuan Chang, and Mu-Yen Chen. 2019. Do virtual reality head-mounted displays make a difference? A comparison of presence and self-efficacy between head-mounted displays and desktop computer-facilitated virtual environments. *Virtual Reality* 23, 4 (2019), 437–446. https://doi.org/10.1007/s10055-018-0376-x

[102] Erica Southgate. 2019. Virtual reality for Deeper Learning: An exemplar from high school science. In *2019 IEEE Conference on Virtual Reality and 3D User Interfaces (VR).* IEEE, 1633–1639. https://doi.org/10.1109/VR.2019.8797841

[103] Marc Steen, Menno Manschot, and Nicole De Koning. 2011. Benefits of co-design in service design projects. *International Journal of Design* 5, 2 (2011).

[104] Michal Takac. 2020. Application of Web-based Immersive Virtual Reality in Mathematics Education. In *2020 21th International Carpathian Control Conference (ICCC).* IEEE, 1–6. https://doi.org/10.1109/ICCC49264.2020.9257276

[105] P Truchly, M Medveckỳ, P Podhradskỳ, and M Vančo. 2018. Virtual reality applications in STEM education. In *2018 16th International Conference on Emerging eLearning Technologies and Applications (ICETA).* IEEE, 597–602. https://doi.org/10.1109/ICETA.2018.8572133

[106] Huawei Tu, Susu Huang, Jiabin Yuan, Xiangshi Ren, and Feng Tian. 2019. Crossing-based selection with virtual reality head-mounted displays. In *Proceedings of the 2019 CHI conference on human factors in computing systems.* 1–14. https://doi.org/10.1145/3290605.3300848

[107] Jarke J Van Wijk. 2006. Views on visualization. *IEEE transactions on visualization and computer graphics* 12, 4 (2006), 421–432. https://doi.org/10.1109/TVCG.2006.80

[108] Silvestro V Veneruso, Lauren S Ferro, Andrea Marrella, Massimo Mecella, and Tiziana Catarci. 2020. CyberVR: an interactive learning experience in virtual reality for cybersecurity related issues. In *Proceedings of the International Conference on Advanced Visual Interfaces.* 1–8. https://doi.org/10.1145/3399715.3399860

[109] Maria Virvou and George Katsionis. 2008. On the usability and likeability of virtual reality games for education: The case of VR-ENGAGE. *Computers & Education* 50, 1 (2008), 154–178. https://doi.org/10.1016/j.compedu.2006.04.004

[110] YuLung Wu, Teyi Chan, BinShyan Jong, and TsongWuu Lin. 2003. A web-based virtual reality physics laboratory. In *Proceedings 3rd IEEE International Conference on Advanced Technologies.* IEEE, 455. https://doi.org/10.1109/ICALT.2003.1215178

[111] Dongsong Zhang, J Leon Zhao, Lina Zhou, and Jay F Nunamaker Jr. 2004. Can e-learning replace classroom learning? *Commun. ACM* 47, 5 (2004), 75–79.

[112] Lei Zhang, Doug A Bowman, and Caroline N Jones. 2019. Exploring effects of interactivity on learning with interactive storytelling in immersive virtual reality. In *2019 11th International Conference on Virtual Worlds and Games for Serious Applications (VS-Games).* IEEE, 1–8. https://doi.org/10.1109/VS-Games.2019.8864531

[113] Jiayan Zhao, Peter LaFemina, Julia Carr, Pejman Sajjadi, Jan Oliver Wallgrün, and Alexander Klippel. 2020. Learning in the field: Comparison of desktop, immersive virtual reality, and actual field trips for place-based STEM education. In *2020 IEEE conference on virtual reality and 3D user interfaces (VR).* IEEE, 893–902. https://doi.org/10.1109/VR46266.2020.00012

[114] Li-Huan Zhu, Long Chen, Shengsheng Yang, Daoming Liu, Jixue Zhang, Xianjin Cheng, and Weisheng Chen. 2013. Embryonic NOTES thoracic sympathectomy for palmar hyperhidrosis: results of a novel technique and comparison with the conventional VATS procedure. *Surgical endoscopy* 27, 11 (2013), 4124–4129.